# Anisotropic in-plane thermal conductivity observed in few-layer black phosphorus

Zhe Luo[1,3], Jesse Maassen[2,3], Yexin Deng[2,3], Yuchen Du[2,3], Richard P. Garrelts[1,3], Mark S. Lundstrom[2,3], Peide D. Ye*[,2,3], and Xianfan Xu*[,1,3]

[1] *School of Mechanical Engineering, Purdue University, West Lafayette, Indiana 47907, United States*

[2] *School of Electrical and Computer Engineering, Purdue University, West Lafayette, Indiana 47907, United States*

[3] *Birck Nanotechnology Center, Purdue University, West Lafayette, Indiana 47907, United States*

* Address correspondence to: yep@purdue.edu (P.D.Y.), xxu@ecn.purdue.edu (X.X.).

ABSTRACT

Black phosphorus has been revisited recently as a new two-dimensional material showing potential applications in electronics and optoelectronics. Here we report the anisotropic in-plane thermal conductivity of suspended few-layer black phosphorus measured by micro-Raman spectroscopy. The armchair and zigzag thermal conductivities are ~20 and ~40 W m$^{-1}$ K$^{-1}$ for black phosphorus films thicker than 15 nm, respectively, and decrease to ~10 and ~20 W m$^{-1}$ K$^{-1}$ as the film thickness is reduced, exhibiting significant anisotropy. The thermal conductivity anisotropic ratio is found to be ~2 for thick black phosphorus films and drops to ~1.5 for the thinnest 9.5-nm-thick film. Theoretical modeling reveals that the observed anisotropy is primarily related to the anisotropic phonon dispersion, whereas the intrinsic phonon scattering rates are found to be similar along the armchair and zigzag directions. Surface scattering in the black phosphorus films is shown to strongly suppress the contribution of long-mean-free-path acoustic phonons.

MAIN TEXT

## Introduction

The successful isolation of graphene[1,2] has inspired rapidly growing research efforts on two-dimensional (2D) materials in the last decade, among which the extensively studied ones include transition metal dichalcogenides (TMDs)[3–5] and hexagonal boron nitride (hBN)[6,7]. 2D materials attract an enormous amount of interest due to their extraordinary electronic, optical, and mechanical properties comparing with bulk counterparts. Recently, black phosphorus (BP) was revisited as a new 2D material with high hole mobility and moderate on/off ratio demonstrated on few-layer BP field-effect transistors (FETs)[8–13]. Unlike semi-metallic graphene with zero band gap[2] and $MoS_2$ with a direct band gap of ~1.8 eV only in its monolayer form[5], BP exhibits a thickness-dependent direct band gap varying from ~0.3 eV (bulk) to > 1.4 eV (monolayer)[9,14–16]. Such tunable band gap benefits few-layer BP in optoelectronic applications such as phototransistors, p-n diodes, and solar cells[10,17–20]. In addition, BP shows intriguing anisotropic properties due to the puckered nature of its in-plane lattice. Several works have explored anisotropic transport in BP, which enables BP for potential applications in novel electronic and optoelectronic devices where the anisotropic properties might be utilized[10,21–23].

Though electronic and photovoltaic properties have been extensively investigated, thermal transport studies of BP, especially experimental ones, are still lacking. Recently the thermoelectric power of bulk BP has been reported, indicating that BP could be used as an efficient thermoelectric material at around 380 K (ref. 24). Some recent first-principles studies also raised interest of BP in thermoelectric applications, claiming that because of the anisotropic lattice structure, the "armchair" direction possesses high electrical conductivity and low lattice thermal conductivity which is desirable for thermoelectrics[21,25–28]. On the contrary, such "orthogonal" electronic and phononic transport of BP may not be favorable in typical FET and photovoltaic devices as low thermal conductivity in the channel direction can lead to thermal management issues. However, to the best of our knowledge, the only thermal conductivity measurement of BP was conducted in 1965 by G. A. Slack on a bulk poly-crystalline BP sample[29], which neither addressed the anisotropic thermal properties of BP nor the thermal transport particularly in few-layer BP.

Here we report the anisotropic in-plane thermal conductivity of suspended few-layer BP measured by micro-Raman spectroscopy, which has been used for measuring thermal conductivity of 2D materials such as graphene[30,31], MoS$_2$[32,33], and hBN[34]. Under laser illumination, the local temperature rise leads to softened atomic bonds and anharmonic phonon coupling as found in graphene[35], inducing a change in the optical phonon frequency and causing the red shift of the corresponding Raman peaks. Such laser-induced temperature-dependent Raman scattering is utilized as an optical thermometer, while the focused laser itself acts as a steady-state heat source. In this work, we produced few-layer BP films of 9.5 to 29.6 nm thickness via the "Scotch tape" method and suspended them on 3-μm-wide slits fabricated on free-standing silicon nitride (SiN) substrate films to measure the in-plane thermal conductivity of BP using the micro-Raman technique. The measured thermal conductivity ranges from ~10 to ~20 W m$^{-1}$ K$^{-1}$ along the armchair direction and ~20 W m$^{-1}$ K$^{-1}$ to ~40 W m$^{-1}$ K$^{-1}$ along the zigzag direction, showing strong thickness-dependence and significant anisotropy in the *x-y* plane. Theoretical calculations of heat transport properties of few-layer BP, based on a first-principles phonon dispersion and a phenomenological treatment of scattering, demonstrate that the anisotropic phonon dispersion along the two directions is responsible for the observed anisotropy, as the scattering rates are found to be nearly isotropic. The thickness dependence of the thermal conductivity is attributed to the strong surface scattering of acoustic phonons, especially phonons with long mean-free-path (MFP).

**Results**

**Sample preparation and polarized-Raman characterization.** Bulk BP consists of puckered honeycomb atomic sheets bonded by van der Waals (vdW) force and therefore can be mechanically exfoliated into atomically thin layers. Figure 1a illustrates its layered lattice structure. BP flakes were tape-exfoliated from bulk crystals and then released onto transparent polymer films for optical examination under microscope. The large, visually uniform and transparent ones were selected as candidates. Before the flakes were transferred, the two principal lattice axes which were generally referred to as the "armchair" and "zigzag" axes, were determined using polarized Raman spectroscopy. To observe the polarized Raman scattering, a linear polarizer was placed at the spectrometer entrance. With the detection polarization perpendicular (referred to as "VH configuration" where "V" stands for vertical laser polarization and "H" for horizontal detection polarization) or parallel (VV configuration) to the incident laser polarization, the optical phonon modes of different symmetries can be selected or eliminated when lattice principal axes are aligned with the laser polarization. In the case of BP, the $A_g$ modes and the $B_{2g}$ mode can be filtered out in VH and VV configurations, respectively, when either the armchair or zigzag axis is aligned with the laser polarization, as shown in Fig. 1b. In this way we were able to identify the armchair or zigzag axis by, for example, observing the $B_{2g}$ mode Raman intensity in the VV configuration while rotating the BP flake. To further distinguish these two axes, we looked into the $A_g^2/A_g^1$ Raman intensity ratio in the VV configuration. The armchair-oriented atomic vibrations of $A_g^2$ phonons lead to maximized $A_g^2$ Raman intensity when laser polarization is along the armchair direction, while the $A_g^1$ Raman intensity remains unchanged because the $A_g^1$ phonon vibrations are out-of-plane[36] (Fig. 1a). Therefore, the $A_g^2/A_g^1$ intensity ratio becomes larger (~2) with armchair-polarized laser excitation, and smaller (~1) with zigzag-polarized laser excitation (Fig. 1b), which serves as Raman signatures of armchair and zigzag lattice axes. The polarized-Raman method is simple yet effective and accurate in determining the lattice orientation of BP crystals. More details of the polarized-Raman measurements can be found in Supplementary Fig. 1 and Supplementary Note 1.

Upon determination of the lattice orientation, the candidate BP flakes were aligned (with angular uncertainty less than ± 3°) and transferred to 3-μm-wide slits fabricated on 200-nm-thick free-standing SiN membranes, details of which can be found in Methods and Supplementary Fig. 2. The rectangular geometry, along with a nearly one-dimensional laser heat source in the center which will be described later, guarantees that the heat conduction is most sensitive to the thermal conductivity perpendicular to the slit. Two slits were patterned to be mutually perpendicular to form a "T" shape, so that the armchair and zigzag thermal transport can be separately investigated on the same flake. Scanning electron microscope (SEM) images and optical images were taken for surface characterization purpose, and multiple atomic force microscope (AFM) scans were performed to obtain the thickness of the flake as well as to confirm the thickness uniformity. Figs. 1c–d are SEM and optical images of a successfully transferred 16.1-nm-thick flake on T-shaped slits. Polarized Raman spectra were also collected on the suspended areas to ensure that the lattice was correctly aligned with the slit. For all subsequent Raman measurements, unpolarized detection was utilized instead polarized detection, in order to achieve maximum signal collection efficiency and reduce the measurement uncertainty.

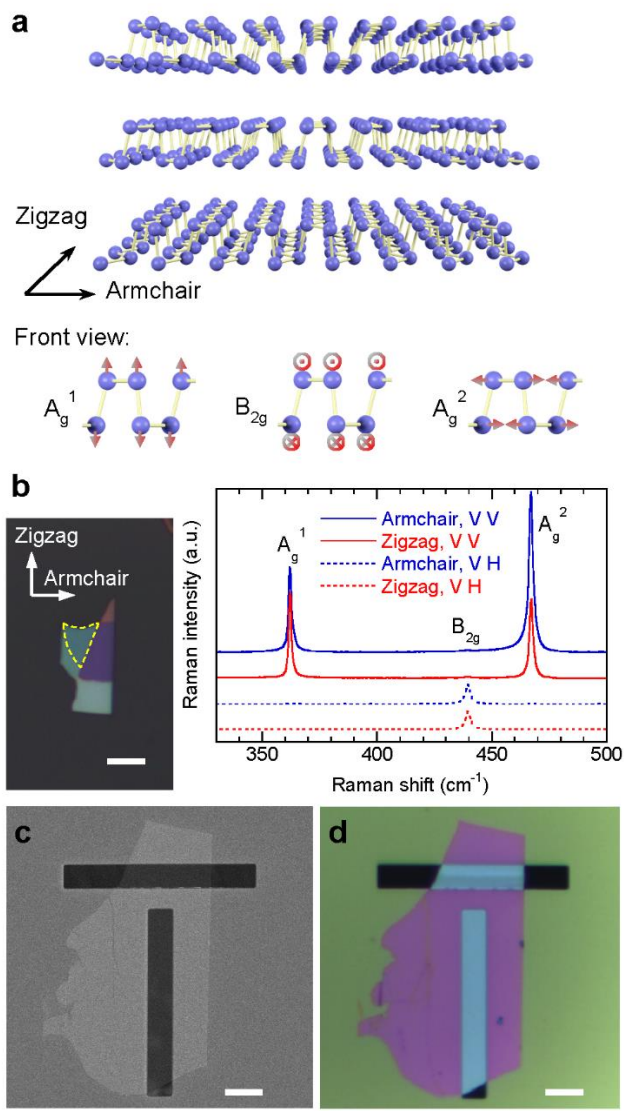

**Figure 1 | Characterization of black phosphorus flakes.** (a) Lattice structure of black phosphorus (BP) and atomic vibrational patterns of $A_g^1$, $B_{2g}$ and $A_g^2$ phonon modes. (b) Polarized Raman spectra (right) collected from a BP flake on $SiO_2$/Si substrate (left) showing the ability of polarized-Raman technique to distinguish the armchair and zigzag axes. (c) Scanning electron microscopy image and (d) optical image of the 16.1-nm-thick BP flake suspended on slits. Scale bars are all 5 μm.

**Thermal measurements by micro-Raman spectroscopy.** The Raman optical thermometer, i.e. the temperature-dependence of Raman scattering, was calibrated using a Raman spectrometer and a heating stage purged with nitrogen. Throughout the calibration, the excitation laser power used was less than 150 μW to minimize excessive heating effect. Taking into consideration of the anisotropy, separate calibrations were performed with both armchair- and zigzag-polarized laser excitations. Figure 2a shows several Raman spectra at different temperatures collected from the 9.5-nm-thick suspended BP film under armchair-polarized excitation, in which the Raman peaks shift towards lower frequency upon heating. In the small-temperature-rise regime, the Raman mode frequency can be expressed as $\omega = \omega_0 + \chi\theta$ with higher-order terms neglected, where $\omega_0$ is the frequency at room temperature, $\theta$ is the temperature rise, and $\chi$ is the temperature coefficient[33]. For the 9.5-nm-thick film, the extracted temperature coefficients are $\chi_{armchair,A_g^1} = -0.01895$ cm$^{-1}$ K$^{-1}$, $\chi_{armchair,B_{2g}} = -0.02434$ cm$^{-1}$ K$^{-1}$, $\chi_{armchair,A_g^2} = -0.02316$ cm$^{-1}$ K$^{-1}$, $\chi_{zigzag,A_g^1} = -0.02175$ cm$^{-1}$ K$^{-1}$, $\chi_{zigzag,B_{2g}} = -0.02877$ cm$^{-1}$ K$^{-1}$ and $\chi_{zigzag,A_g^2} = -0.02700$ cm$^{-1}$ K$^{-1}$, which are in agreement with previously reported values[37]. It is seen that the zigzag-polarized excitation yielded temperature coefficients of larger absolute values (see Fig. 2b for $A_g^2$ mode as an example), which might be caused by anisotropic thermal expansion during the heating process. Among the three Raman modes, the $A_g^2$ mode was found most sensitive to temperature change as well as showing highest Raman intensity, therefore was chosen as the thermometer. Raman shift is also sensitive to strain and stress, aside from temperature change. The evaluation of such effects is provided in Supplementary Note 2.

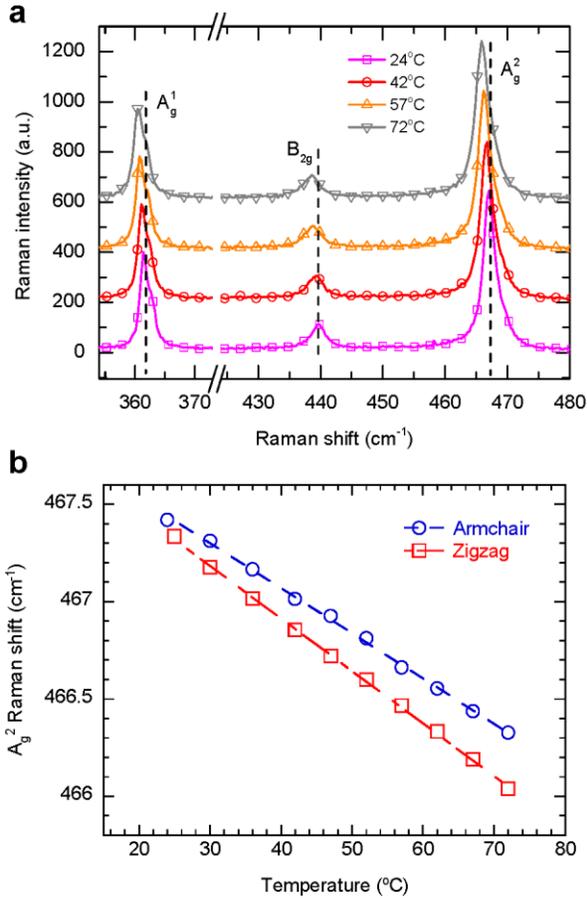

**Figure 2 | Raman thermometer calibration results of the 9.5-nm-thick black phosphorus film.** (a) Four sample Raman spectra taken at 24, 42, 57 and 72°C with armchair-polarized laser. The dashed lines correspond to the peak positions at 24°C. (b) The $A_g^2$ Raman shift as a function of temperature for both armchair- and zigzag- polarized laser. The dashed lines show linear fit results.

Micro-Raman thermal conductivity experiments were conducted at room temperature in nitrogen atmosphere. Figure 3a illustrates the experimental setup. To best achieve the desired one-dimensional heat transfer, a 75-μm-wide rectangular aperture was placed in front of the objective lens to produce a laser focal line instead of a circular spot. In the direction where the aperture cuts the laser beam, the partially filled objective lens aperture produces a larger width at the focal point, yielding a stretched line-shaped focal spot. The Gaussian width $w_0$ and length $l_0$ (both defined as the radius where the intensity drops to $1/e^2$) of the laser focal line is characterized to be 0.39 μm and 3.1 μm, respectively (Fig. 3b). The laser focal line was aligned to the centerline of the slit, creating a one-dimensional heat source at the center of the suspended BP film. Consequently, the dominant heat flow occurred perpendicular to the slit, enabling the isolation of heat conduction along the armchair or zigzag direction.

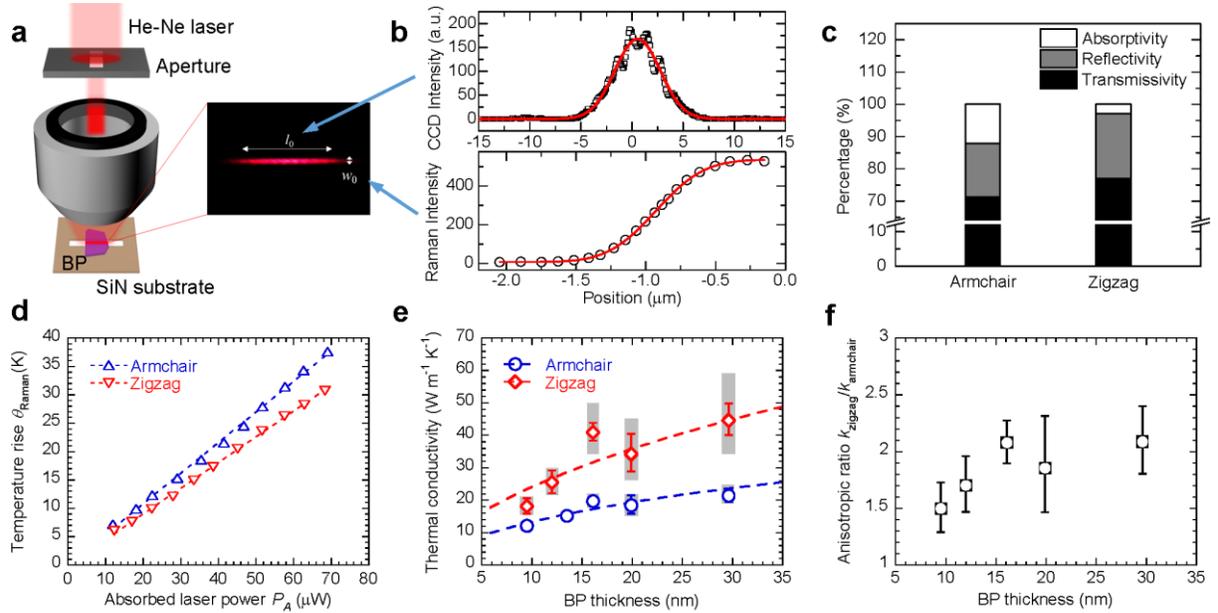

**Figure 3 | Thermal conductivity measurements of black phosphorus using micro-Raman technique.** (a) Illustration of the experimental setup and an optical image of the produced laser focal line. (b) The lengthwise profile, and the knife-edge-measured widthwise integrated profile of the laser focal line. The solid lines are Gaussian function and error function curve fits, respectively. (c) The optical absorptivity $A$, reflectivity $R$, and transmissivity $T$ of the 9.5-nm-thick suspended black phosphorus (BP) film upon armchair- and zigzag-polarized laser incidence. (d) Laser-power-dependent temperature rise ($\theta_{Raman}$) of the 16.1-nm-thick BP film determined by the micro-Raman spectroscopy along armchair and zigzag transport directions. The dashed lines are linear fits. (e) Extracted armchair and zigzag in-plane thermal conductivities ($k_{armchair}$ and $k_{zigzag}$) of multiple BP films. Dashed lines are results of theoretical modeling. The gray error bars account for the uncertainty of SiN substrate thermal conductivity $k_{SiN}$, while the blue/red error bars do not. (f) The anisotropic ratio $k_{zigzag}/k_{armchair}$ at different BP thicknesses. The ratio at 12-nm thickness is calculated using linearly interpolated armchair thermal conductivity from adjacent thicknesses.

At various incident laser powers, a series of Raman spectra were collected and converted to the local temperature rise at the laser focal line using the previously obtained temperature coefficient. Figure 3d plots the Raman-measured temperature rise $\theta_{Raman}$ versus the absorbed laser power $P_A$ for the 16.1-nm-thick suspended BP film, in which the data shows a linear correlation similar to other micro-Raman experiments[30–34]. The absorbed laser power was determined by $P_A = AP =$ $(1-R-T)P$, where $P$ is the incident laser power, $A$ is the absorptivity, $R$ is the reflectivity , and $T$ is the transmissivity. The reflectivity of BP films is measured using a beam splitter in the incident laser path, which deviates the reflected light to a separate path where its intensity is measured; by comparing the reflected light intensity of the BP films and a silver-coated mirror as a reference, the reflectivity of BP films can be calculated. The transmissivity of the BP films is measured under the slit, by dividing the transmitted laser intensity on the BP-covered slit by that at the adjacent empty slit. Note that $A$, $R$ and $T$ are all anisotropic quantities (Fig. 3c) due to anisotropic optical conductivity, and our measured absorptivity of the 9.5-nm-thick suspended film (~12.0% for armchair polarization and ~2.9% for zigzag polarization) agrees well with the theoretical predictions[22]. Due to the much higher absorption of armchair-polarized light in BP films, all the thermal conductivity measurements were carried out with an armchair-polarized laser beam in order to reduce the uncertainty of $P_A$. The uncertainty of the absorptivity of the BP films we measured ranges from 0.2–0.7%.

**Extracting the in-plane thermal conductivity of BP.** Three-dimensional anisotropic heat conduction equation

$$\frac{\partial}{\partial x}\left(k_x\frac{\partial\theta}{\partial x}\right)+\frac{\partial}{\partial y}\left(k_y\frac{\partial\theta}{\partial y}\right)+\frac{\partial}{\partial z}\left(k_z\frac{\partial\theta}{\partial z}\right)+\dot{q}=0 \tag{1}$$

was solved using the finite volume method, with the heat source term for stretched laser focal line given as

$$\dot{q}=\frac{\alpha\left(1-R\right)P}{\pi w_0 l_0}e^{-\left(w^2/w_0^2+l^2/l_0^2\right)}e^{-\alpha z}, \tag{2}$$

where $\alpha$ is the absorption coefficient, $w$ and $l$ represent the coordinates corresponding to the width and length directions of the focal line. The temperature rise within the laser focal line is expressed in Gaussian-weighted-average form

$$\theta_{\text{Raman}}=\frac{\int_0^t\int_0^\infty\int_0^\infty\theta e^{-\left(w^2/w_0^2+l^2/l_0^2+2\alpha z\right)}\mathrm{d}w\,\mathrm{d}l\,\mathrm{d}z}{\int_0^t\int_0^\infty\int_0^\infty e^{-\left(w^2/w_0^2+l^2/l_0^2+2\alpha z\right)}\mathrm{d}w\,\mathrm{d}l\,\mathrm{d}z}, \tag{3}$$

where $t$ is the BP film thickness, and the factor 2 comes from the absorption of Raman scattered photons within the film when they are traveling backwards to the surface. It is shown that theoretically $\theta_{\text{Raman}}$ varies linearly with absorbed laser power $P_A$ for thin films[38]. By comparing the experimentally measured slopes $\mathrm{d}\theta_{\text{Raman}}/\mathrm{d}P_A$ with the calculated slopes, one can extract both the armchair and zigzag thermal conductivity ($k_{armchair}$ and $k_{zigzag}$) of the suspended BP film by iterative calculations between the two directions (Supplementary Note 3), since even though the temperature is predominately affected by the thermal conductivity along the direction perpendicular to the slit, it is still weakly affected by the thermal conductivity along the other direction. It is also worth noting that, though the in-plane thermal transport dominates in the suspended region, the cross-plane conduction may need to be accounted for in the supported region as it directly affects the heat sink efficiency. Therefore, we evaluated the cross-plane heat conduction in the supported region and performed separate micro-Raman and time-domain thermoreflectance (TDTR) measurements of the thermal conductivity of SiN substrate, details of which are included in Supplementary Fig. 3 and Supplementary Note 4.

## Discussion

The measured anisotropic in-plane thermal conductivity values of few-layer BP are summarized in Fig. 3e. The electronic contribution to the total thermal conductivity is estimated to be a small fraction given typical carrier concentrations in few-layer BP (Supplementary Note 5), therefore the data presented here can be largely attributed to the lattice. The measured $k_{zigzag}$ is generally 50–100% higher than $k_{armchair}$, showing strong anisotropic feature[21]. It arises mostly from the anisotropic phonon dispersion along Γ–X (armchair) and Γ–A (zigzag) directions as discussed further below. In Fig. 3e, $k_{zigzag}$ starts from ~40 W m$^{-1}$ K$^{-1}$ for thick films over 15 nm, then sharply decreases to ~20 W m$^{-1}$ K$^{-1}$; similar trend can also be found for $k_{armchair}$, decreasing from ~20 W m$^{-1}$ K$^{-1}$ to ~10 W m$^{-1}$ K$^{-1}$. This strong thickness dependence originates from significant surface scattering of long-MFP phonons; similar results were observed in few-quintuple-layer $Bi_2Te_3$ films[39], where surface scattering was found to heavily affect electron and phonon transport. The anisotropic trend of larger thermal conductivity along zigzag compared to armchair direction is consistent with previous theoretical calculations[25,26,40]. The values of 110 W m$^{-1}$ K$^{-1}$ (zigzag) and 36 W m$^{-1}$ K$^{-1}$ (armchair) predicted through first-principles modeling by Jain and McGaughey[26] are larger compared to those reported in this work. Other *ab initio* calculations[25,40] provided thermal conductivities closer to our experimental values, but these smaller values were shown to result from approximations in the theoretical approach[26]. The predicted thermal conductivities were obtained assuming perfect monolayer BP, however in the presence of surface roughness or adsorbates, greater phonon scattering can occur leading to smaller thermal conductivities, which is the main source of discrepancy between the previous predictions and the measured values reported here. The measured thermal conductivity of few-layer BP is comparable with the reported value of bulk BP (12.1 W m$^{-1}$ K$^{-1}$)[29]; however the bulk BP was poly-crystalline so that this thermal conductivity value was merely an averaged value along three lattice axes, among which the cross-plane axis exhibits lower thermal conductivity due to the weak vdW interlayer bonding. Figure 3f presents the thermal conductivity anisotropic ratio $k_{zigzag}/k_{armchair}$, which drops from ~2 to ~1.5 as BP thickness decreases to less than 10 nm. Note that the error bars in Fig. 3f did not account for the uncertainty of $k_{SiN}$, since any variation in SiN substrate thermal conductivity will affect both $k_{armchair}$ and $k_{zigzag}$, which cannot be treated by standard error propagation method as it requires uncertainties to be mutually independent.

Theoretical modeling of few-layer BP, based on *ab initio* phonon dispersion calculations and phenomenological scattering models, was conducted to understand the experimental results. The phonon dispersion of bulk BP (valid for our thin films, see Supplementary Fig. 4 and Supplementary Note 6) along the high-symmetry reciprocal lattice points (Fig. 4a) was calculated with the optimized lattice constants given as $a = 4.57$ Å, $b = 3.30$ Å and $c = 11.33$ Å (ref. 41), and the results are shown in Fig. 4b. The use of given lattice constants is justified by reproducing electron and phonon dispersion of bulk BP, details of which are included in Supplementary Figs 5–6 and Supplementary Note 7. There are 3 acoustic branches and 9 optical branches extending up to roughly 55 meV, with an energy gap between 35 meV and 40 meV. Significant differences of acoustic phonon bandwidths and group velocities are observed between Γ–X (armchair) and Γ–A (zigzag) (and also Γ–Z cross-plane) directions, which are responsible for the measured thermal conductivity anisotropy. We then computed the transport properties solving the phonon Boltzmann equation using the Landauer approach[39,42], which expresses the thermal conductivity as

$$k = K_0 \int M_{ph}(\varepsilon)\, \lambda_{ph}(\varepsilon,T) W_{ph}(\varepsilon,T)\, \mathrm{d}\varepsilon \,, \tag{4}$$

where $K_0 = \pi^2 k_B^2 T/3h$ is the quantum of thermal conductance, $M_{ph}$ is the number of conducting modes per cross-sectional-area (units: m$^{-2}$), $\lambda_{ph}$ is the phonon MFP for backscattering which includes Umklapp phonon-phonon scattering and surface scattering, and $W_{ph}(\varepsilon,T) = (3\varepsilon/\pi^2 k_B^2 T)[\partial n_{BE}(\varepsilon,T)/\partial T]$ is a normalized "window function" with $n_{BE}$ being the Bose-Einstein distribution and $\varepsilon$ the phonon energy. It can be seen that the above formula of thermal conductivity consists of two quantities: the number of phonon modes per cross-sectional-area $M_{ph}$ which depends only on the phonon dispersion, and the phonon MFP $\lambda_{ph}$ which depends on both the phonon dispersion and the scattering mechanisms. $M_{ph}$ can be readily and efficiently calculated using the "band-counting" algorithm, as implemented in the simulation tool LanTraP[43]. The calculated phonon modes $M_{ph}$ along the armchair and zigzag directions are presented in Fig. 4c, where $M_{ph}$ along the zigzag direction shows a significantly larger number of modes, particularly for the acoustic phonons in the energy range 5–15 meV, which typically carry most of the heat. This anisotropy in number of modes is related to the larger phonon group velocity along the zigzag direction compared to the armchair direction, as observed from the slopes of the phonon dispersion. At a given temperature, the average number of thermally active phonon modes is given by[42]

$$\langle M_{ph} \rangle = \int M_{ph}(\varepsilon) W_{ph}(\varepsilon) \mathrm{d}\varepsilon \ . \tag{5}$$

In the absence of scattering, the ballistic thermal conductance $K^{ball}$, which is independent of sample length, is simply $\langle M_{ph} \rangle$ times the quantum of thermal conductance $K_0$. The calculated temperature-dependent $\langle M_{ph} \rangle$ and $K^{ball}$ are shown in Figs 4d–e, and they display a strong anisotropy between zigzag and armchair directions (note that the calculated phonon dispersion is extrapolated to other temperatures, since we assume it does not significantly change in the plotted temperature range). Above 150 K, their respective zigzag-to-armchair ratio is almost constant and equal to ~1.5. Given that the thermal conductivity $k$ can be written as $K^{ball}$ times the average phonon MFP $\langle \lambda_{ph} \rangle = \int \lambda_{ph}(\varepsilon) M_{ph}(\varepsilon) W_{ph}(\varepsilon) \mathrm{d}\varepsilon \big/ \int M_{ph}(\varepsilon) W_{ph}(\varepsilon) \mathrm{d}\varepsilon$ (here we define an average phonon MFP to help explain our results; it is important to note that the full MFP distribution is used in our theoretical calculations, which can span orders of magnitude), the thermal conductivity ratio $k_{zigzag}/k_{armchair}$ can be expressed as $\langle M_{ph} \rangle_{zigzag} \big/ \langle M_{ph} \rangle_{armchair}$ times $\langle \lambda_{ph} \rangle_{zigzag} \big/ \langle \lambda_{ph} \rangle_{armchair}$. This suggests that the ~1.5–2.0 anisotropic ratio of measured thermal conductivity is in large part due to $\langle M_{ph} \rangle_{zigzag} \big/ \langle M_{ph} \rangle_{armchair}$, a quantity that depends only on the phonon dispersion and shows a zigzag-to-armchair ratio of ~1.5 at room temperature. As the BP film thickness decreases, the minor contribution of $\langle \lambda_{ph} \rangle_{zigzag} \big/ \langle \lambda_{ph} \rangle_{armchair}$ (i.e. the anisotropy in phonon MFP) further decreases due to the enhanced surface scattering. Thus the thermal conductivity anisotropic ratio $k_{zigzag}/k_{armchair}$ decreases and approaches ~1.5, as depicted in Fig. 3f.

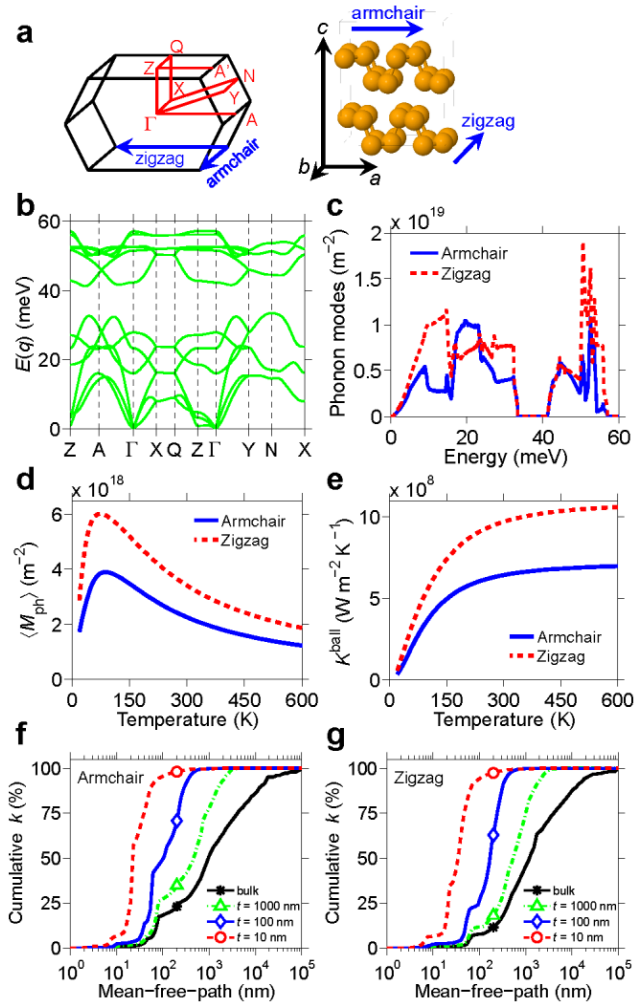

**Figure 4 | First-principles-based modeling results of few-layer black phosphorus.** (a) High symmetry points in the Brillouin zone (left) and crystal structure (right) of black phosphorus. (b) Phonon dispersion (energy $E$ versus momentum $q$) along high symmetry points. (c) Number of conducting phonon modes per cross-sectional-area versus energy. (d) Average number of thermally active phonon modes per cross-sectional-area as a function of temperature. (e) Ballistic thermal conductance as a function of temperature. (f)–(g) Normalized cumulative thermal conductivity at 300 K versus phonon mean-free-path for backscattering (using phenomenological scattering models for Umklapp and surface scattering with specularity parameter $p = 0$) for transport along armchair and zigzag directions.

To calculate the in-plane thermal conductivity of few-layer BP films, the phonon MFP distribution $\lambda_{ph}(\varepsilon,T)$ in equation (4) must be specified. We included intrinsic Umklapp phonon-phonon scattering and surface scattering of phonons at film boundaries to determine the phonon MFP distribution. Each scattering mechanism, Umklapp and surface scattering, has one free parameter that can be adjusted to fit the measured thermal conductivity (see Supplementary Note 7 for further details). Using the Fuchs-Sondheimer[39,42] approach to include the effect of surface scattering in few-layer BP, the strength of the surface scattering is controlled through a specularity parameter $p$, which dictates the degree to which scattering at the surface is specular (denotes an atomically smooth surface where momentum along the transport direction is conserved) or diffusive (denotes a rough surface where momentum along the transport direction is randomized). We note that this surface scattering model was originally developed assuming isotropic bulk scattering, and it was adopted in this work to investigate anisotropic BP as the best available candidate. Our theoretical model provides a good match to the measured thermal conductivity with a specularity parameter $p$ in the range 0–0.4 (Supplementary Fig. 7), shown as dashed curves in Fig. 3e (with $p = 0$). This is consistent with the fact that few-layer BP is susceptible to chemical reactions with oxygen and moisture[14] as well as possible minor PMMA residue (even undetectable with Raman spectroscopy and AFM), which may have enhanced surface scattering. Interestingly, the intrinsic Umklapp scattering rate distribution used to calculate $k$ in Fig. 3e is the same along both armchair and zigzag directions. With an isotropic scattering rate, the intrinsic phonon MFP is longer along the zigzag direction compared to the armchair direction (i.e. anisotropic), due to the difference in phonon group velocities. Even with isotropic MFP, the thermal conductivity anisotropy is prominent (Supplementary Fig. 8). Our theoretical analysis suggests that the measured thermal conductivity anisotropy is primarily a consequence of the particular phonon dispersion of BP, as opposed to strong anisotropic scattering of phonons. Previous first-principles ballistic calculations of monolayer BP, which only consider phonon dispersion, also display significant anisotropy in the thermal transport properties[27].

To further illustrate how phonons of different MFP contribute to the thermal conductivity and the impact of surface scattering, in Figs 4f–g we plot the normalized cumulative $k$ versus MFP at different BP film thicknesses. They were obtained using the first-principles-calculated phonon dispersion and the phenomenological scattering models which were calibrated to the measured BP film thermal conductivities, as described above. It can be clearly seen how surface scattering strongly alters the contribution of long-MFP phonons. For example, in a 10-nm-thick film, nearly 90% of the heat is carried by phonons with MFP less than 100 nm, as opposed to roughly 20% and 10% in bulk along armchair and zigzag directions, respectively. Please note that the plots of cumulative $k$ versus MFP for thicknesses greater than 30 nm are an extrapolation. For bulk BP, a significant contribution to $k$ comes from phonons with MFP around 1 μm, while for 10-nm-thick BP film the largest contribution comes from phonons with MFP near 30 nm. Therefore, for thermoelectric applications where higher electro-thermal conversion efficiency results from lower $k$, further reductions in few-layer BP thermal conductivity could be achieved by selectively scattering phonons that contribute the most[44]. However in cases where a large thermal conductivity is beneficial (e.g. to mitigate hot-spots in electronic devices), thicker BP films are desirable because of less surface scattering. Alternatively, finding solutions to enhance surface smoothness and reduce diffusive surface scattering, such as encapsulating BP by an inert hBN layer[45], could be equally effective.

In summary, we experimentally measured the anisotropic in-plane thermal conductivity of suspended few-layer BP films using micro-Raman spectroscopy and performed first-principles-based theoretical studies to gain insight into the anisotropic thermal transport. The armchair and zigzag thermal conductivities, $k_{armchair}$ and $k_{zigzag}$, are ~20 W m$^{-1}$ K$^{-1}$ and ~40 W m$^{-1}$ K$^{-1}$ for BP films over 15 nm thickness, respectively. Strong surface scattering of phonons reduces $k_{armchair}$ and $k_{zigzag}$ by half, down to ~10 W m$^{-1}$ K$^{-1}$ and ~20 W m$^{-1}$ K$^{-1}$, respectively, for the thinnest 9.5-nm-thick BP film. The thermal conductivity anisotropic ratio $k_{zigzag}/k_{armchair}$ is found to be ~2 for thick BP films and drops to ~1.5 for the thinnest one. Theoretical modeling of thermal transport in few-layer BP reveals that the observed anisotropic thermal conductivity stems from the material's particular phonon dispersion, as opposed to anisotropic scattering. Diffusive surface scattering is found to be prevalent in few-layer BP, which strongly reduces the contribution of long-MFP phonons. Our results may provide useful guidance for the design of BP-based thermoelectric, electronic, and optoelectronic devices from the perspective of anisotropic thermal transport.

During the preparation of this paper, we became aware of recently published angle-dependent Raman spectroscopy works on BP which exploited the same polarized-Raman technique[46,47].

**Methods**

**Flake preparation and transfer.** On the Si wafer, ~1-μm-thick poly(vinyl alcohol) (PVA) was spin-coated and baked at 70°C for 5 minutes and ~200-nm-thick poly(methyl methacrylate) (PMMA, 950 A4) was spin-coated onto PVA and baked using the same recipe. Silicon dicing tape was used to peel flakes off the bulk BP crystal, then the flakes were released on to the PMMA/PVA stack. The polymer stack was then cleaved off, flipped over and mounted on a glass frame for subsequent visual examination under microscope. Once a candidate flake was found, polarized-Raman measurements were conducted to identify the lattice orientation. Then the flake along with the PMMA/PVA was aligned in desirable direction and attached to the slits fabricated by focused-ion-beam (FEI Nova) on the 200-nm-thick free-standing SiN substrate film. The whole sample was then dipped into acetone to dissolve the PMMA layer and to remove the PVA layer, then finally dried with nitrogen. Supplementary Figure 2 illustrates the process. We used large amount of acetone (> 70 mL) and very long soaking time (> 12 hours) to minimize the PMMA residue on the BP films, and the surface cleanliness was confirmed by AFM scans and Raman spectra. The BP flakes were not baked or annealed throughout the preparation and transfer process to avoid excessive oxidation and to retain the crystallinity. During the flake preparation and transfer process, the BP flake was exposed to the air for about 1 hour, while all the subsequent Raman measurements were performed in dry nitrogen atmosphere.

**Micro-Raman and SEM/AFM measurements.** A HORIBA LabRAM HR800 Raman spectrometer equipped with a 632.8 nm wavelength He-Ne laser and an Olympus 100× objective lens was used for all the Raman measurements. The nominal spectral resolution is 0.27 cm$^{-1}$ per pixel, and the Lorentzian peak fitting yields a peak position shift uncertainty less than 0.02 cm$^{-1}$, corresponding to a temperature rise measurement uncertainty of less than 1 K for BP. A linear polarizer (Thorlabs, LPNIRE050-B) was used for polarized-Raman measurements. The SEM images were taken by an FEI Nova system and the AFM measurements were carried out on an AIST-NT CombiScope system.

**Laser focal line characterization.** Along the length direction, the intensity profile was extracted directly from the raw RGB data of a CCD camera and fitted with Gaussian function to obtain $l_0$

$$I_l = C_l e^{-(l-l_c)^2/l_0^2}. \tag{6}$$

We performed knife-edge measurements to characterize the width of the focal line. A Si wafer with sharp edge was gradually brought into the focal line along width direction by a piezoelectric stage. As the stage moved, the Raman intensity of Si increased as more Si was exposed to the laser. Similar to the length direction, the intensity profile along the width direction can also be described by a Gaussian function

$$I_w = C_w e^{-(w-w_c)^2/w_0^2}, \tag{7}$$

and the Raman intensity of Si as a function of stage position is the integration of $I_w$ along width direction

$$I_{w,\text{Raman}}(x) = \int_{-\infty}^{x} I_w \mathrm{d}w = \int_{-\infty}^{x} C_w e^{-(w-w_c)^2/w_0^2} \mathrm{d}w = \frac{\sqrt{\pi}}{2} C_w w_0 \left[\operatorname{erf}(x) + 1\right]. \tag{8}$$

By fitting the measured intensity into the above equation we were able to obtain $w_0$.

ACKNOWLEDGEMENTS

This work was supported in part by DARPA MESO (Grant No. N66001-11-1-4107) and through the NCN-NEEDS program, which is funded by the National Science Foundation, Contract No. 1227020-EEC, ECCS-1449270, ARO W911NF-14-1-0572 and the Semiconductor Research Corporation. J.M. acknowledges financial support from NSERC of Canada.

AUTHOR CONTRIBUTIONS

X.X. and P.D.Y. conceived the idea, designed and supervised the experiments. Z.L. performed the majority of the experiments and analyzed the experimental data. J.M. and M.S.L. conducted the theoretical calculations and analysis. Y.D., Y.D. and Z.L. performed the polarized-Raman experiments. R.P.G. performed the TDTR measurements. Z.L. and J.M. co-wrote the manuscript with input from all authors.

COMPETING FINANCIAL INTERESTS STATEMENT

The authors declare no competing financial interests.

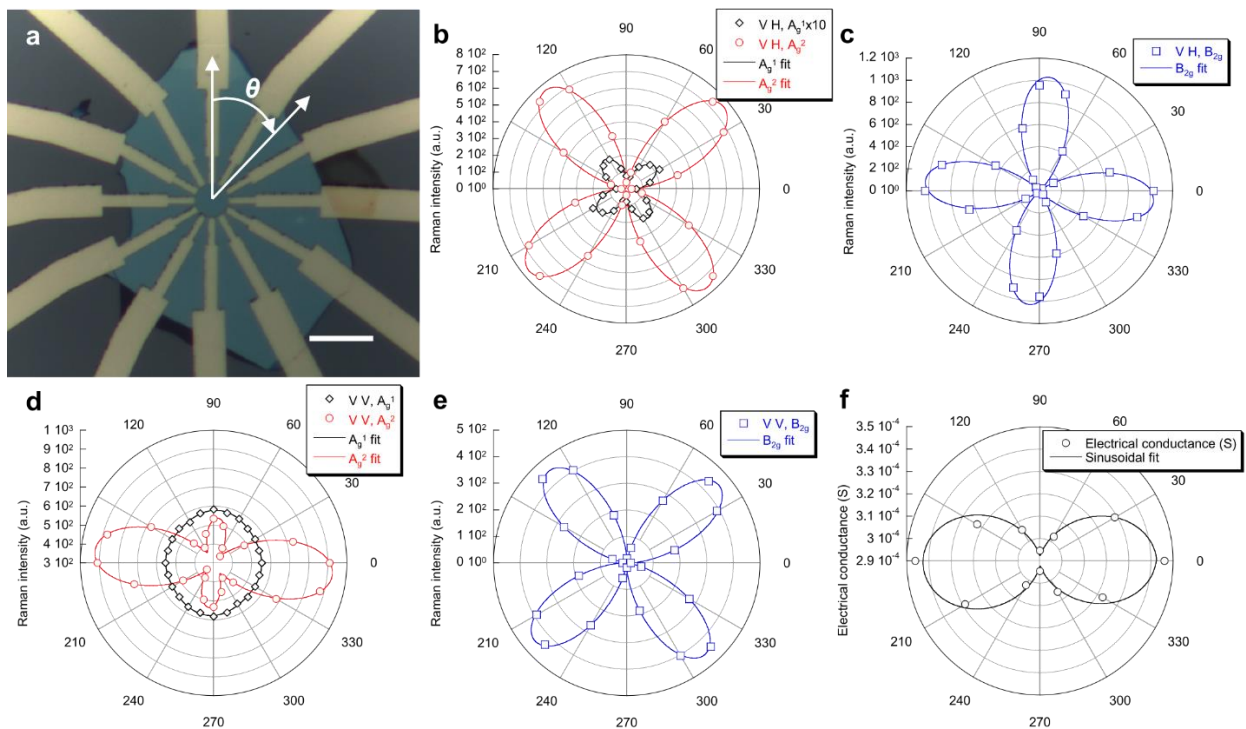

**Supplementary Figure 1 | Polarized-Raman and electrical conductance measurements on a 32-nm-thick black phosphorus (BP) flake.** (a) Optical image showing the flake and the electrodes. Scale bar is 10 μm. Angle-resolved polarized Raman intensity of (b) $A_g$ modes (c) $B_{2g}$ mode in VH configuration, and (d) $A_g$ modes (e) $B_{2g}$ mode in VV configuration. In (b) to (e), solid lines are all curve fits based on Supplementary Equation 3. (f) Angle-resolved electrical conductance measured by six pairs of electrodes using 50 mV voltage. Solid line is the sinusoidal curve fit.

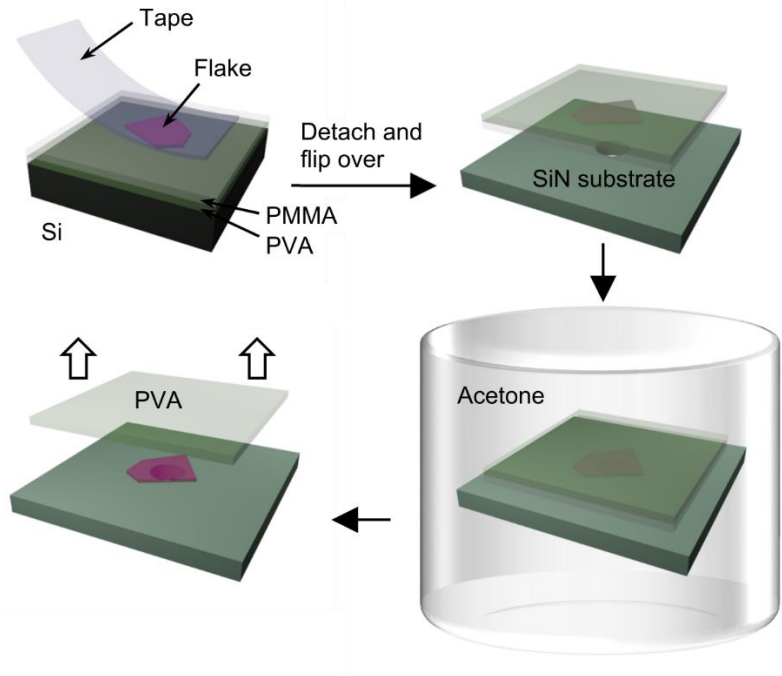

**Supplementary Figure 2 | Sketch of the flake preparation and transfer process.**

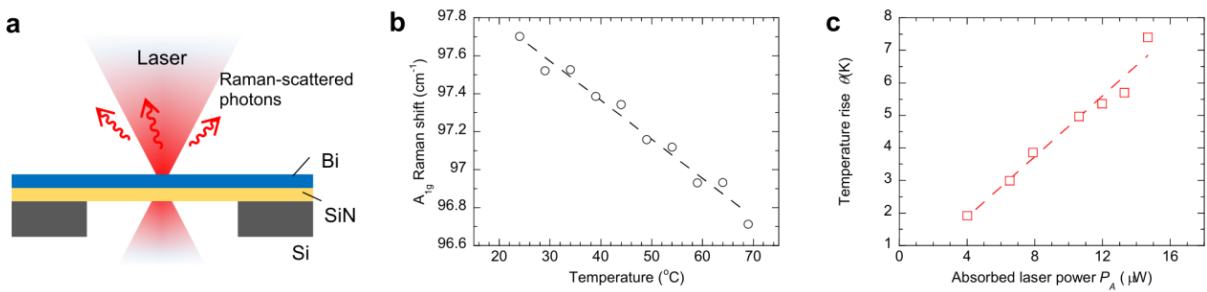

**Supplementary Figure 3 | Micro-Raman measurement of the Bi-SiN sample.** (a) Sketch of the sample structure and experimental setup. (b) $A_{1g}$ Raman thermometer calibration results. Dashed line is the linear fit showing temperature coefficient $\chi_{A_{1g}} = -0.0206$ cm$^{-1}$. (c) Laser-power-dependent temperature rise, and the linear fit with slope $d\theta/dP_A = 0.4671$ K μW$^{-1}$. The Raman-measured temperature uncertainty is ~1 K.

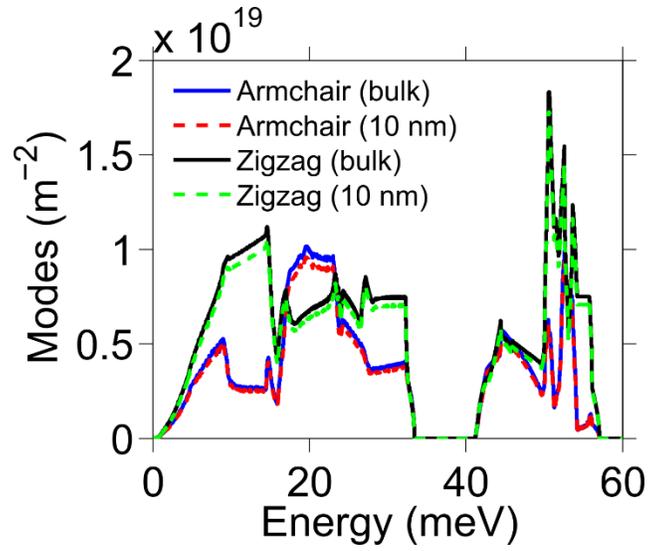

**Supplementary Figure 4 | Impact of confinement on number of phonon modes in BP.** Number of modes per cross-sectional-area $M_{ph}$ versus phonon energy for bulk BP and a 10-nm-thick BP film. The number of modes in the film was calculated assuming that phonons with half-wavelength (in the cross-plane direction) greater than the film thickness do not contribute to the number of modes.

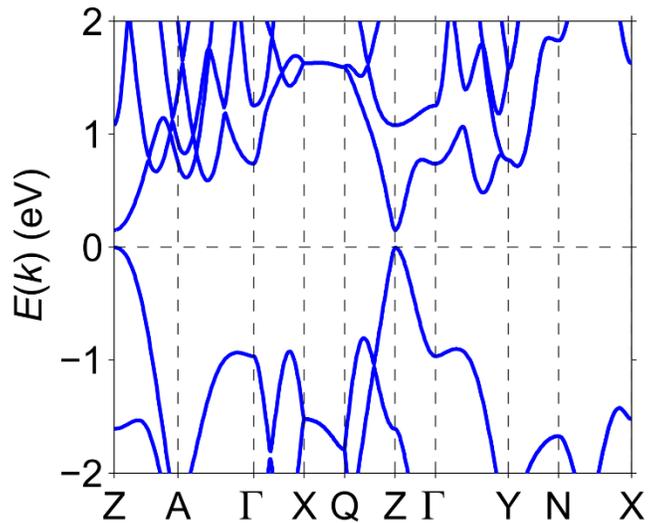

**Supplementary Figure 5 | Electron dispersion of bulk BP.** Electron band structure of bulk BP along high symmetry points.

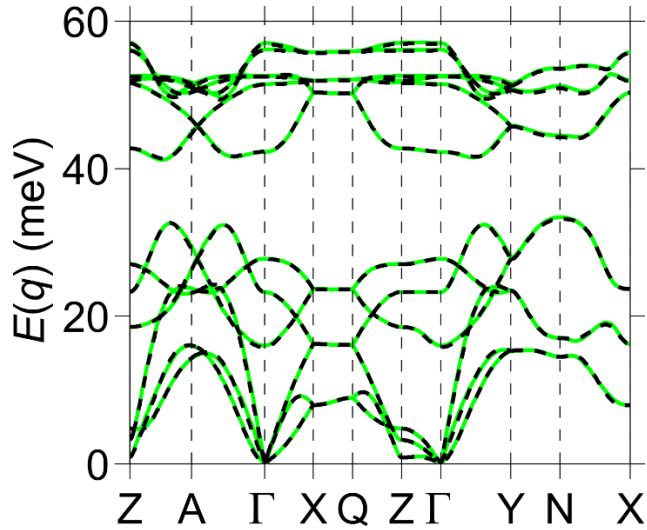

**Supplementary Figure 6 | Phonon dispersion of bulk BP.** Comparison of phonon dispersion along high-symmetry points using the lattice constants reported in *Nat. Commun.* **5**, 4475 (2014) $a = 4.57$ Å, $b = 3.30$ Å, $c = 11.33$ Å (solid black lines) and our total-energy minimized lattice constants $a = 4.56$ Å, $b = 3.31$ Å, $c = 11.32$ Å (dashed green lines).

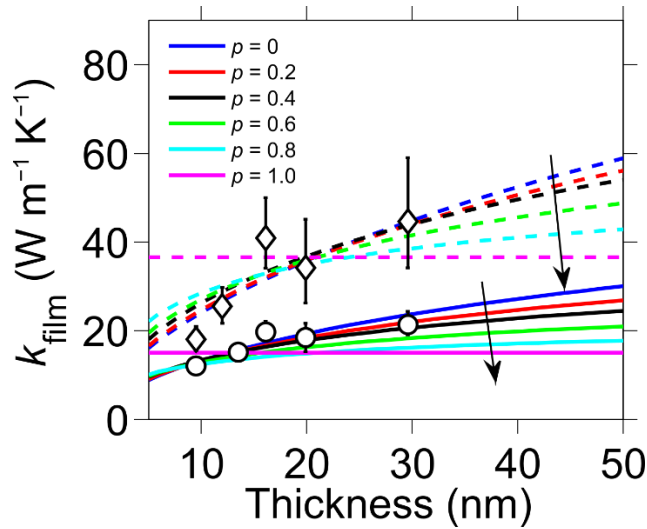

**Supplementary Figure 7 | Effect of specularity parameter $p$.** Thermal conductivity versus BP film thickness for zigzag (diamonds, dashed lines) and armchair (circles, solid lines) directions, as a function of specularity parameter $p$. For a given $p$ value, the $B$ parameter controlling the strength of the Umklapp scattering is adjusted to best match the measured values (symbols). For $p$ ranging from 0 to 1 with 0.2 increments, the optimal $B$ was found to be $2 \times 10^{-19}$ s K$^{-1}$, $3 \times 10^{-19}$ s K$^{-1}$, $4.2 \times 10^{-19}$ s K$^{-1}$, $6.3 \times 10^{-19}$ s K$^{-1}$, $9.5 \times 10^{-19}$ s K$^{-1}$, $1.6 \times 10^{-18}$ s K$^{-1}$, respectively. The arrows indicate the trend for increasing $p$.

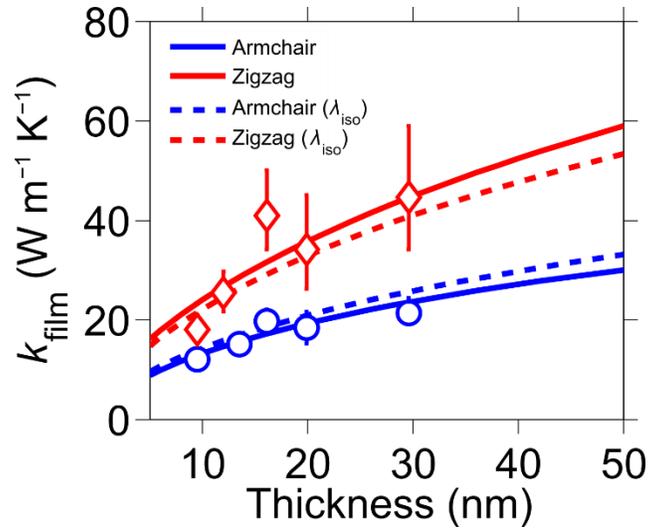

**Supplementary Figure 8 | Anisotropic versus isotropic phonon mean-free-path (MFP).**
Thermal conductivity versus BP film thickness along armchair and zigzag directions. Markers are experimental data. Solid lines are calculated using an anisotropic MFP distribution (i.e. isotropic scattering time distribution), as described by our methods. Dashed lines are calculated using an isotropic MFP distribution obtained by averaging the anisotropic MFP over armchair and zigzag directions at each energy. In the case of an isotropic MFP, significant anisotropy in thermal conductivity is observed arising from the anisotropy in the phonon modes.

**Supplementary Note 1 | Polarized-Raman measurements**

The Raman scattering intensity $I$ is proportional to[1]

$$I \propto \left| \mathbf{e}_i \times \mathbf{R} \times \mathbf{e}_s \right|^2 ,$$ (1)

where $\mathbf{e}_i$, $\mathbf{e}_s$ represent incident and scattered light polarizations, respectively, and $\mathbf{R}$ is the Raman tensor of a specific Raman mode. In BP lattice $abc$ coordinates where $a$ is the zigzag direction, $b$ is the cross-plane direction and $c$ is the armchair direction, the incident laser polarization is expressed as $\mathbf{e}_i = (\cos\theta, 0, \sin\theta)$ where $\theta$ is the angle between the $a$ axis and the laser polarization vector (Supplementary Fig. 1a), and the detection polarization $\mathbf{e}_s$ is set as $\mathbf{e}_s \perp \mathbf{e}_i$ (VH configuration) or $\mathbf{e}_s \| \mathbf{e}_i$ (VV configuration). For BP, three Raman-active modes $A_g^1$ $B_{2g}$ and $A_g^2$ can be observed, and their Raman tensors are[2]

$$\mathbf{R}_{A_g} = \begin{pmatrix} A & & \\ & B & \\ & & C \end{pmatrix}, \quad \mathbf{R}_{B_{2g}} = \begin{pmatrix} & & E \\ & & \\ E & & \end{pmatrix},$$ (2)

then the Raman intensity can be written as

$$\text{VH:} \begin{cases} I_{A_g} \propto \dfrac{(C-A)^2}{4} \sin^2 2\theta \\ I_{B_{2g}} \propto E^2 \cos^2 2\theta \end{cases} \quad \text{VV:} \begin{cases} I_{A_g} \propto \left( A\cos^2\theta + C\sin^2\theta \right)^2 \\ I_{B_{2g}} \propto E^2 \sin^2 2\theta \end{cases}.$$ (3)

When incident laser polarization is along the zigzag direction ($\theta = 0°$),

$$\text{VH:} \begin{cases} I_{A_g} = 0 \\ I_{B_{2g}} \propto E^2 \end{cases} \quad \text{VV:} \begin{cases} I_{A_g} \propto C^2 \\ I_{B_{2g}} = 0 \end{cases},$$ (4)

and when incident laser polarization is along the armchair direction ($\theta = 90°$),

$$\text{VH:} \begin{cases} I_{A_g} = 0 \\ I_{B_{2g}} \propto E^2 \end{cases} \quad \text{VV:} \begin{cases} I_{A_g} \propto A^2 \\ I_{B_{2g}} = 0 \end{cases}.$$ (5)

It can be clearly seen that $I_{A_g}$ in VV configuration can be used to distinguish zigzag and armchair axes. Supplementary Figure 1 presents collected polarized Raman intensity from a 32-nm-thick BP flake and sinusoidal fitting using Supplementary Equation 3, showing clear angle-dependent signature. Note that for $A_g^1$ mode Raman tensor, $A = C$, so that only $A_g^2$ mode can be used to determine the lattice orientation (armchair where intensity gets maximized) as shown in Supplementary Fig. 1d. To further confirm the lattice orientation, six pairs of 0.3 nm Ti/40 nm

Pd/40 nm Au electrodes were patterned with 30° interval on the flake using electron beam lithography to measure angle-dependent electrical conductivity, and the results are shown in Supplementary Fig. 1f. It has been demonstrated that electrical conductance is higher along the armchair direction[3], so that our electrical measurement validates the lattice orientation derived from polarized-Raman measurements.

## Supplementary Note 2 | Excessive strain and stress effects on the Raman-based thermal measurements of BP films

Raman shift is known to be sensitive to strain and stress. In our micro-Raman experiments, the heating-induced thermal strain contributes to a part of the temperature coefficient and is calibrated during the calibration processes. The only undesirable strain or stress comes from the thermal expansion mismatch between the BP film and the SiN membrane during heating. Since the thermal expansion of BP $\alpha_{BP}$ ~$50 \times 10^{-6}$ $K^{-1}$ (ref. 4) is larger than that of SiN $\alpha_{SiN}$ ~$2 \times 10^{-6}$ $K^{-1}$ (ref. 5), a compressive stress will possibly be applied to the BP film due to the constraint of the SiN membrane. However, due to the large aspect ratio of the BP film ($< 30$ nm thickness vs. 3 μm width), the compressive stress is expected to be relaxed by natural buckling and thus does not affect our Raman measurements.

To further evaluate this stress effect, we conducted separate calibrations on both suspended regions (two perpendicular slits) of the ~20-nm-thick BP film. The compressive stress on the two regions, if any, would be perpendicular to the slits and therefore orthogonal to each other. Since the thermal expansion and the Raman-strain response of BP are anisotropic[6], the orthogonal compressive stress would lead to different temperature coefficient if such stress plays a significant role. The obtained temperature coefficients on the two regions are $-0.0260$ $cm^{-1}$ $K^{-1}$ and $-0.0268$ $cm^{-1}$ $K^{-1}$, respectively, with only a minute relative difference of ~3%. This indicates that the suspended regions of BP films remain free from excessive strain and stress that originate from the constraint of SiN membrane during heating, thus we believe that the compressive stress due to thermal expansion mismatch has a negligible effect on our Raman measurements.

It is worth noting that there are other Raman indicators such as peak linewidth and anti-Stokes to Stokes intensity ratio which can serve as temperature probes and are insensitive to strain and stress; however, they are not as accurate as the Stokes peak shift[7,8]. Trade-offs between stress-sensitivity and temperature-accuracy should be carefully considered, based on the instrument capability and samples to be investigated. If Raman peak shift thermometer is to be implemented, the strain and stress effects should be carefully evaluated.

## Supplementary Note 3 | Determination of $k_{armchair}$ and $k_{zigzag}$ and their uncertainties

Although the heat transfer in suspended BP films is close to 1D, $k_{armchair}$ and $k_{zigzag}$ are still weakly coupled. Therefore we extract them using the iterative calculations using our numerical heat transfer model. For a given flake, it is suspended on two perpendicular slits for measuring $k_{zigzag}$ and $k_{armchair}$ separately, which are referred to as ZZ slit and AC slit here for convenience. First we assumed a $k_{zigzag}$ and applied it to the modeling of the AC slit to extract a $k_{armchair}$; then the extracted $k_{armchair}$ was in turn applied to the modeling of the ZZ slit and a new $k_{zigzag}$ was obtained. Then the previously assumed $k_{zigzag}$ was replaced by this new value, and the above process was iteratively repeated until both $k_{armchair}$ and $k_{zigzag}$ converge.

The uncertainties of $k_{armchair}$ and $k_{zigzag}$ are determined using the standard error propagation method, i.e. $\sigma_k = [\Sigma(A_i\sigma_i)^2]^{1/2}$ where $\sigma_i$ is the uncertainty of a single parameter and $A_i$ represents the linearized effect this parameter has on the final result. We have included all uncertainties of concern such as that of data slope $d\theta/dP$, temperature coefficient $\chi$, thickness $t$, absorptivity $A$, and SiN thermal conductivity $k_{SiN}$. Since the relative uncertainties of $d\theta/dP$, $\chi$, $t$ and $A$ are the same or approximately the same for the two directions for a given flake, the relative uncertainty of $k$, which is $\sigma_k/k$, remains approximately the same as well, according to error propagation theory. This yields $\sigma_{k,zigzag}/\sigma_{k,armchair}$ ~ $k_{zigzag}/k_{armchair}$ ~1.5–2, which explains the larger colored error bars of $k_{zigzag}$ in Fig. 3e. For gray error bars, we took into consideration the uncertainty of $k_{SiN}$. It has larger impact on the uncertainty of $k_{zigzag}$ than $k_{armchair}$ as seen in Fig. 3e. The reason is that, in zigzag measurements, the temperature distribution across the slit is more flat due to the higher thermal conductivity in the zigzag direction, therefore the temperature is higher at the edge of the slit so that the conductivity of the heat sink (SiN) plays a bigger role than the armchair case. This means that $\sigma_{k,\text{zigzag}}$ is more sensitive to $\sigma_{k,\text{SiN}}$ than $\sigma_{k,\text{armchair}}$, which explains the even longer gray error bars of $k_{\text{zigzag}}$.

**Supplementary Note 4 | Evaluation of the convective heat transfer loss, the heat sink efficiency, and measurement of SiN thermal conductivity $k_{\text{SiN}}$**

To minimize the convective heat transfer loss due to the nitrogen flow, the flow rate is controlled at a minimum level of ~7 cm$^3$ s$^{-1}$. Considering the cross-sectional area of a few cm$^2$ of our measurement chamber, the nitrogen flow speed at the sample surface would be on the order of 1 cm s$^{-1}$, so that the effect of the convection was close to that of free convection. To further quantify the convection effect, we performed numerical analysis on the thinnest 9.5 nm flake which would likely show largest convective heat transfer loss due to its largest surface-to-volume ratio. Typical free convective heat transfer coefficient ranges from 2–25 W m$^{-2}$ K$^{-1}$ (ref. 9). We calculated $h =$ 0 W m$^{-2}$ K$^{-1}$ and $h =$ W m$^{-2}$ K$^{-1}$ cases and found the armchair (zigzag) thermal conductivity to be 12.14 (18.22) W m$^{-1}$ K$^{-1}$ and 11.99 (17.94) W m$^{-1}$ K$^{-1}$, respectively. The relative difference due to convective heat transfer loss is only ~1.5%. $h = 20$ W m$^{-2}$ K$^{-1}$ was implemented in our modeling as the best estimate, which would lead to an even smaller difference in the thermal conductivity values.

We also analyzed the heat conduction through a 200-μm-thick layer of N$_2$ between the BP film and the sample holder and the heat conduction in N$_2$ from the top of the BP film to the objective lens through a distance of 210-μm, the working distance of the objective lens. Numerical calculations on the 9.5-nm-thick BP film found that the extracted armchair and zigzag thermal conductivity to be 11.77 and 17.42 W m$^{-1}$ K$^{-1}$, respectively. Comparing with the case considering convection only with h = 20 W m$^{-2}$ K$^{-1}$, in which the armchair and zigzag thermal conductivity are 12.06 and 18.08 W m$^{-1}$ K$^{-1}$ respectively, the relative difference is less than 3.8%. Note that for thicker BP films this discrepancy will be smaller, since the heat conduction within BP films will become more prominent.

To assess the effects of cross-plane heat conduction in the supported region of the BP films, we performed sensitivity studies on three key parameters: the cross-plane thermal conductivity of BP $k_{z,BP}$, the contact thermal resistance at BP/SiN interface $R_c$ and the thermal conductivity of SiN membrane $k_{SiN}$. It turns out that $k_{z,BP}$ and $R_c$ do not affect much the fitted BP thermal conductivity. Varying $k_{z,BP}$ from 0.5 to 5 W m$^{-1}$ K$^{-1}$, which is considered typical for vdW bonds[10], only resulted in ~1% difference of the extracted BP thermal conductivity; varying $R_c$ from $1 \times 10^{-9}$ to $1 \times 10^{-7}$ m$^2$ K W$^{-1}$, typical for vdW interfaces[11], gave ~4.5% difference. In contrast, $k_{SiN}$ has a larger impact, as a variation from 8 to 10 W m$^{-1}$ K$^{-1}$ (intermediate value in literature reported range[12–14]) leads to as much as 50% change of the fitted BP thermal conductivity. Hence, we conducted separate micro-Raman measurements to measure $k_{SiN}$ using thermally evaporated bismuth (Bi) film as the Raman transducer, which has been validated by the successful measurements of amorphous Al$_2$O$_3$ thin films[15].

Bi (99.999% purity, Sigma Aldrich) was thermally evaporated onto fresh 200-nm-thick SiN membrane (Product# 21520, Ted Pella) at the vacuum level of ~1×10$^{-6}$ Torr. A clean glass slide was coated simultaneously, and the Bi film on glass was scratched intentionally so that the Bi film thickness could be determined using AFM. The thickness was measured to be 34.7 ± 1.3 nm. Then the same micro-Raman technique, except that the focal spot was circular instead of line-shaped, was utilized on the Bi-SiN film stack as sketched in Supplementary Fig. 3a. The Bi-SiN sample was calibrated (Supplementary Fig. 3b) and measured (Supplementary Fig. 3c) using A$_{1g}$ mode (~97 cm$^{-1}$) as the Raman thermometer. The slope d$\theta$/d$P_A$ was fitted into a 2D radial heat transfer model where the thermal conductivity of Bi $k_{Bi}$ was taken from our previous work (8.4 ± 0.9 W m$^{-1}$ K$^{-1}$)[15]. Note that the Bi films were prepared under the same condition, so that the use of previous obtained $k_{Bi}$ was considered valid. By fitting the model, $k_{SiN}$ is extracted to be 9.4 + 1.3/−1.1 W m$^{-1}$ K$^{-1}$ which is used for extracting the thermal conductivity of BP, along with $k_{z,BP}$ taken as 1 W m$^{-1}$ K$^{-1}$ and $R_c$ taken as $2 \times 10^8$ m$^2$ K W$^{-1}$. The absorptivity of the Bi-SiN sample was measured in the same way as described in the "Results" Section of the main text. The uncertainty of measured $k_{SiN}$ was determined using the error propagation method (see Supplementary Note 3) and propagated into the uncertainty of measured BP thermal conductivity. It is noted that $k_{SiN}$ reported in the literature ranges from 1–13 W m$^{-1}$ K$^{-1}$, which we found might be related to the composition of the SiN films[12,14,16–18]. With Si composition increasing, $k_{SiN}$ increases and reaches > 8 W m$^{-1}$ K$^{-1}$ for Si:N = 1:0.99 (ref. 14). The Si:N atomic ratio of the SiN films used in this study is a rather high 1:0.92, thus justifying the $k_{SiN}$ measured to be 9.4 W m$^{-1}$ K$^{-1}$ which falls into the higher side of the reported range.

We also attempted to measure $k_{SiN}$ using the time-domain thermoreflectance (TDTR) method. A 150-nm-thick Au layer was coated on the SiN sample, so that the film stack structure from top to bottom is: 150 nm Au, 200 nm SiN, 240 nm SiO$_2$, bulk (200 μm) Si. A standard TDTR setup equipped with a 532-nm Nd:YAG pulsed laser (~5 ns pulsewidth, 5 kHz repetition rate) as the pump laser and a 633-nm He-Ne continuous wave laser as the probe laser was used to obtain the data. The free parameters were $k_{SiN}$, and the contact resistances of the Au-SiN ($R_1$), SiN-SiO$_2$ ($R_2$), and SiO$_2$-Si ($R_3$) interfaces. All the other parameters were taken from widely used bulk values. However, because the contact resistance values are unknown, $k_{SiN}$ was not conclusively determined. With all contact resistances varied by only ~20%, we were able to fit the experimental data with both $k_{SiN} = 5$ W m$^{-1}$ K$^{-1}$ and 10 W m$^{-1}$ K$^{-1}$.

## Supplementary Note 5 | Estimating the electronic contribution to the thermal conductivity

To estimate the electronic thermal conductivity of BP, we again used the Landauer approach[31]. By following essentially the same procedure outlined above for phonons, we extracted the electron dispersion from DFT (the scissor method was employed to adjust the band gap to the experimentally accepted value 0.33 eV), computed the number of modes per cross-sectional-area for electrons, in which a constant electron MFP was used. The equations relating number of modes and MFP to electrical conductivity, Seebeck coefficient, and electronic thermal conductivity are provided in ref. 22. In ref. 3, the authors measured the hole mobility of a 8-nm-thick BP film at room temperature to be ~420 cm$^2$ V$^{-1}$ s$^{-1}$ and ~270 cm$^2$ V$^{-1}$ s$^{-1}$ along the armchair and zigzag directions, respectively. Using the measured hole concentration of ~$1.7 \times 10^{19}$ cm$^{-3}$ combined with our calculation of carrier concentration from the density-of-states, we determined that the Fermi level is located at 0.06 eV into the valence band. By fitting the MFP for electrons along armchair and zigzag to reproduce the mobility values, we found $\lambda_{armchair} = 13$ nm and $\lambda_{zigzag} = 38$ nm. From this we can calculate the electronic thermal conductivity: ~0.65 W m$^{-1}$ K$^{-1}$ (armchair) and ~0.45 W m$^{-1}$ K$^{-1}$ (zigzag). The lattice contribution to the thermal conductivity is significantly larger than the electronic contribution, and the latter can be neglected in the analysis of the measured thermal conductivity.

**Supplementary Note 6 | Impact of confinement on the phonon dispersion**

In the theoretical calculations we assumed that the phonon dispersion of bulk BP represents well the phonon states in the experimental few-layer BP films, since calculating the phonon dispersion for each film thickness would be prohibitively demanding. In order to validate this assumption, we computed the number of modes per cross-sectional-area $M_{ph}$ assuming that all phonons with half-wavelength greater than the film thickness do not contribute to the number of modes to estimate the impact of confinement. With this calculation, we can estimate the proportion of phonons which have wavelengths similar to the thickness therefore would likely be affected by the confinement. Supplementary Figure 4 compares $M_{ph}$ versus energy for bulk BP and a 10-nm-thick film. Only minor differences are observed, justifying our model using the phonon dispersion of bulk BP.

**Supplementary Note 7 | Calculation of lattice thermal conductivity**

We computed the thermal transport properties of few-layer BP using the Landauer approach, which has been successfully used to describe phonon transport in thin films[19,20]. Within this approach the lattice thermal conductivity $k$ is expressed in equation (4) in the main article, where we can see that the two main quantities needed to obtain the thermal conductivity are the number of modes per cross-sectional-area $M_{ph}$ and the phonon MFP $\lambda_{ph}$ (referred to as $\lambda_{film}$ below in order to distinguish from the phonon MFP of bulk BP).

$M_{ph}$ is calculated using equation (A4) in ref. 21 or equivalently equation (16) in ref. 22, which depends only on phonon dispersion. The phonon dispersion of bulk BP (Fig. 4b in the main article) is calculated by first relaxing the atomic coordinates to achieve forces less than 0.001 eV Å$^{-1}$, using the optimized lattice constants provided in the main article ($a$ = 4.57 Å, $b$ = 3.30 Å and $c$ = 11.33 Å). Our lattice constants, taken from ref. 23, are consistent with other calculated values for monolayer and bilayer BP[24,25], and are reasonably close to the experimental lattice constants[26] ($a$ = 4.38 Å, $b$ = 3.31 Å and $c$ = 10.48 Å). We perform density functional theory (DFT) simulations using GGA-PBE for exchange-correlation potential and PAW method to capture the effect of the atomic core, as implemented in VASP[27,28]. We employ a plane-wave energy cutoff of 450 eV and a 11×9×9 Monkhorst-Pack-generated reciprocal-space grid (lattice vectors: $a_1$ = [$a$ 0 0]; $a_2$ = [0 $b/2 -c/2$]; $a_3 = [0 \ b/2 \ c/2]$). Supplementary Figure 5 shows our calculated electronic dispersion of bulk BP, which is consistent with that reported in ref. 23, including a direct band gap of 0.15 eV. The dynamical matrix is constructed using the force constants extracted from the finite-displacement method (displacement of 0.01 Å, 3×5×5 supercell of the primitive cell with first number corresponding to $a$-axis), and then solved to obtain the phonon energies using Phonopy[29]. We used the phonon energies, calculated for a rectangular cell ($a×b×c$) with a 85×117×35 $k$-point grid in reciprocal space, to extract the number of modes per cross-sectional-area $M_{ph}$ using LanTraP[30]. Although our computational approach is the same as that in ref. 23 (from which we take the lattice constants), our own total-energy minimization yields slightly different optimal lattice constants $a = 4.56$ Å, $b = 3.31$ Å and $c = 11.32$ Å. Supplementary Figure 6 compares the phonon dispersion of bulk BP using the lattice constants reported in ref. 23 and those obtained through our optimization. No significant differences in phonon energy are observed, thus we can safely use the lattice constants provided in ref. 23.

The phonon mean-free-path (MFP) for backscattering includes two contributions: *i)* phonon-phonon (Umklapp) scattering and *ii)* surface scattering of phonons on the finite thickness of the BP films. The intrinsic (bulk) MFP is written as[19]

$$\lambda_{bulk}\left(E,T\right) = 2\left(\frac{\int_{BZ} v_{\parallel}^{+}\left(E,k_{\perp}\right) M_{ph}\left(E,k_{\perp}\right) \mathrm{d}k_{\perp}}{\int_{BZ} M_{ph}\left(E,k_{\perp}\right) \mathrm{d}k_{\perp}}\right)\tau_{U}\left(E,T\right), \tag{6}$$

where $v_{\parallel}^{+}$ is the band velocity along the transport direction for the forward moving states, $M_{ph}\left(E\right) = \int_{BZ} M_{ph}\left(E,k_{\perp}\right)\mathrm{d}k_{\perp}$, and $\tau_{U}\left(E,T\right)$ is intrinsic scattering time for Umklapp phonon-phonon scattering. The term in the parentheses depends only on phonon dispersion. Thus, the only missing information to calculate the thermal conductivity is the scattering time. We use the following phenomenological model of Umklapp scattering[21]

$$\tau_{U}^{-1}\left(E,T\right) = B E^2 T \exp\left(-C/T\right)\big/\hbar^2, \tag{7}$$

where $B$ and $C$ are adjustable parameters. We choose $C = 0$ since we focus only on room temperature calculations and wish to eliminate one adjustable parameter. The film MFP is obtained by including surface scattering using the Fuchs-Sondheimer approach[19,20]

$$\lambda_{\text{film}}\left(E\right)=\lambda_{\text{bulk}}\left(E\right)\left[1-\frac{3\left(1-p\right)}{2\delta}\int_1^\infty\left(\frac{1}{x^3}-\frac{1}{x^5}\right)\frac{1-\exp\left(-\delta x\right)}{1-p\exp\left(-\delta x\right)}\mathrm{d}x\right], \tag{8}$$

where $\delta=(4/3)(t/\lambda_{\text{bulk}})$, $t$ is the BP film thickness and $p$ is the specularity parameter controlling the degree of resistive scattering at the surface, with $p=0$ and $1$ corresponding to completely diffuse and specular scattering, respectively. To summarize, we adjust the two parameters $B$ (Umklapp scattering) and $p$ (surface scattering) to achieve the best agreement with the measured thermal conductivities (note that these parameters do not change depending on the transport direction, zigzag versus armchair). For $p < 1$, it can be shown that as the film thickness $t$ is reduced, $\lambda_{\text{film}}$ tends to be more isotropic along armchair and zigzag directions due to the increasing effect of surface scattering over the intrinsic Umklapp scattering.

With the number of modes per cross-sectional-area $M_{\text{ph}}$ and the phonon MFP $\lambda_{\text{ph}}$ determined, we were able to calculate the thermal conductivity of BP films and found good agreement with the experimental data using $B = 2 \times 10^{-19}$ s K$^{-1}$ and $p = 0$ (completely diffusive surface scattering).

Although BP has anisotropic thermal transport characteristics, here we adopted a well-established and physics-based model for an isotropic scattering time distribution due to Umklapp scattering, which describes the temperature and energy dependence. This model was found to provide good agreement with experiment (see for example ref. 21). Since we are unaware of well-proven phenomenological models for Umklapp scattering time distributions that are anisotropic, we began by trying a widely-used and successful isotropic scattering model. Our results show that an isotropic scattering time distribution adequately reproduces the experimental thermal conductivity; there is no clear signature in the experimental data of an anisotropic Umklapp scattering rate (i.e. the observed anisotropy in the thermal conductivity is mostly the result of anisotropy in the modes). It is possible that any anisotropy in the Umklapp scattering time is not clearly perceived in the experimental data since surface scattering tends to make the MFP distribution more isotropic (i.e. long-MFP phonons are scattered more strongly than short-MFP phonons).

To examine what role the anisotropic MFP distribution (i.e. using an isotropic scattering time distribution, as described above) plays in the observed anisotropy in thermal conductivity, we compute the thermal conductivity using an isotropic MFP distribution that corresponds to the value averaged over both zigzag and armchair directions (i.e. angle-averaged MFP) at each energy.

Supplementary Figure 8 compares the calculated thermal conductivity using both anisotropic and isotropic MFP distributions. The isotropic MFP still yields significant anisotropy in thermal conductivity, although less than using an anisotropic MFP. The anisotropy in thermal conductivity is mostly a direct consequence of the anisotropy in the number of phonon modes.

## Supplementary References